\documentclass[12pt,preprint]{aastex}

\slugcomment{Accepted for publication in ApJL}
\shorttitle{Unification of 3CR sources}
\shortauthors{Podigachoski et al.}

\begin{document}

\title{The unification of powerful quasars and radio galaxies \\ and their relation to other massive galaxies}

\author{Pece Podigachoski and Peter Barthel}
\affil{Kapteyn Astronomical Institute, University of Groningen, 9747 AD Groningen, The Netherlands}
\email{podigachoski@astro.rug.nl}
\author{Martin Haas}
\affil{Astronomisches Institut, Ruhr Universit\"{a}t, D-44801 Bochum, Germany}
\author{Christian Leipski}
\affil{Max-Planck Institut f\"{u}r Astronomie (MPIA), D-69117 Heidelberg, Germany}
\and
\author{Belinda Wilkes}
\affil{Harvard-Smithsonian Center for Astrophysics, Cambridge, MA 02138, USA}

\begin{abstract}
The unification model for powerful radio galaxies and radio-loud quasars postulates 
that these objects are intrinsically the same but viewed along different angles. 
\textit{Herschel Space Observatory}\footnote{\textit{Herschel} is an ESA space observatory
with science instruments provided by European-led Principal Investigator consortia 
and with important participation from NASA.} data permit the assessment of that model 
in the far-infrared spectral window. We analyze photometry from \textit{Spitzer} and 
\textit{Herschel} for the distant 3CR hosts, and find that radio galaxies and 
quasars have different mid-infrared, but indistinguishable far-infrared colors. Both 
these properties, the former being orientation dependent and the latter orientation 
invariant, are in line with expectations from the unification model. Adding powerful 
radio-quiet active galaxies and typical massive star-forming galaxies to the analysis, 
we demonstrate that infrared colors not only provide an orientation indicator, but can 
also distinguish active from star-forming galaxies.
\end{abstract}


\section{Introduction}
Active Galactic Nuclei (AGN) show a large variety of observational properties 
despite being universally driven by accretion of matter onto supermassive 
black holes. Their ubiquitous nature and complex interplay with their host 
galaxies make AGN essential elements of any model of galaxy evolution. Since 
their discovery in the 1950s, AGN have been observationally classified into a 
number of different types depending on the particular wavelength range used 
for the observations. Unification models for 
AGN \citep{Barthel89,Antonucci93,Urry&Padovani95} were developed later, 
explaining the differences between some AGN types solely in terms of viewing 
angle. An essential part of these models is the so-called AGN 'torus': a 
geometrically and optically thick structure filled with molecular gas and dust 
surrounding the active nucleus. The AGN torus has a geometry that allows 
radiation to escape freely in certain directions but not in others, rendering 
samples of objects selected at X-ray, ultra-violet (UV) and visible wavelengths 
largely inapplicable for testing unification models.

Radio-selected AGN are particularly well-suited for testing the unification 
model, because when selected based on their extended (well outside the torus), 
low-frequency, transparent (i.e. optically thin) radio emission they show no bias 
with respect to orientation. The unification model for powerful radio-loud AGN 
postulates that radio galaxies (RGs) are viewed edge-on so that the torus obscures 
the active nucleus and nearby broad-line emitting regions, while quasars (QSRs) are 
the same objects, but viewed with a face-on torus so that the nuclear regions and 
broad emission lines are directly visible. The validity of the unification model for 
powerful radio-loud AGN has been examined using a variety of orientation-dependent 
properties \citep[recently reviewed in][]{Antonucci12}, in the radio 
\citep{Barthel89,Singal93}, mid-infrared (MIR) \citep{Ogle06,Cleary07,Haas08,Leipski10}, 
and X-ray \citep{Wilkes13} domains.
\section{Infrared properties}
If the unification model holds, then any isotropic (orientation invariant) property 
must be comparable for samples of RGs and QSRs with matched low-frequency radio 
properties. One such an isotropic candidate is the far-infrared (FIR) emission, which 
is dominated by star-formation-heated dust on the scale of the host galaxy, but may 
also include a contribution from optically thin, AGN-powered emission from the torus. 
Due to sensitivity issues, earlier FIR studies relied on relatively small samples of 
mostly nearby, less luminous sources, and therefore failed to provide robust 
conclusions on the unification model for powerful radio-loud AGN in the FIR 
\citep[e.g.,][]{Hes95,vanBemmel00,Meisenheimer01,Haas04}. 
The \textit{Herschel Space Observatory} \citep{Pilbratt10}, with its unprecedented 
sensitivity, resolution, and wavelength coverage for the first time permits rest-frame 
FIR tests of unification for distant radio-loud AGN which have hitherto been outside 
the reach of infrared space missions. In order to test the unification model in the 
FIR, we study the complete $z>1$ 3CR sample of powerful 
(L$_{\mathrm{178MHz}}$ $>$ 5 $\times$ 10$^{28}$ W Hz$^{-1}$) radio-loud AGN 
\citep{Spinrad85} using the two \textit{Herschel} imaging photometers: PACS at 70 and 
160~$\mu$m \citep{Poglitsch10} and SPIRE at 250, 350, and 500~$\mu$m \citep{Griffin10}. 
The extreme luminosities of our sample objects ensure that all objects are highly 
accreting \citep[in quasar-mode, e.g.,][]{Best&Heckman12}; none of them is classified as 
a low-ionization emission-line source \citep[e.g.,][]{Ogle06}. 

In addition to the \textit{Herschel} data, the availability of ancillary \textit{Spitzer} 
\citep{Werner04} data at wavelengths between 3.6 and 24~$\mu$m \citep{Haas08} allows us to 
obtain physical properties of the 3CR hosts by fitting their full infrared spectral energy 
distributions (SEDs) with a combination of four components, as explained in detail in 
\citet{Barthel12} and \citet{Podigachoski15}. These components include emission from the 
circumnuclear, AGN-heated torus dust \citep{Hoenig&Kishimoto10}, from the extended 
star-formation-heated cold dust, from the evolved stellar populations, and from hot nuclear 
graphite dust. The best-fit SEDs of our 61 3CR hosts were recently presented in 
\citet{Podigachoski15}. Twenty-three hosts in the sample (presented in Table~\ref{tab1}) 
have good signal-to-noise detections in at least three \textit{Herschel} bands, typically 
the three shortest bands, resulting in estimated star formation rates (SFRs) of order 
hundreds of solar masses per year, coeval with the black hole activity \citep{Podigachoski15}. 
Such prodigious SFRs, at the level of ultra-luminous infrared galaxies, have similarly been 
inferred in other studies of high-$z$ radio galaxies \citep[e.g.][]{Ogle12,Drouart14,Tadhunter14}. 
\section{MIR/FIR colors -- a powerful tool}
A simpler but revealing way to look at the shapes of SEDs is to employ colors. Infrared 
color-color diagrams are widely used to separate AGN from non-AGN in wide-/deep-field 
galaxy surveys \citep[e.g.,][]{Lacy04,Stern05,Donley12,Kirkpatrick13}. Using the best-fit SEDs 
of the 23 3CR objects detected in at least three \textit{Herschel} bands, we create 
rest-frame color-color diagrams from their hosts' dust emission at 5, 20, 70, and 100~$\mu$m. 
The emission at the former two wavelengths is mainly AGN-powered \citep[e.g.,][]{Rowan-Robinson95}, 
whereas the emission at the latter two wavelengths is predominantly star-formation-powered 
\citep[][]{Schweitzer06}.  

\begin{figure}
 \epsscale{1.0} 
 \plotone{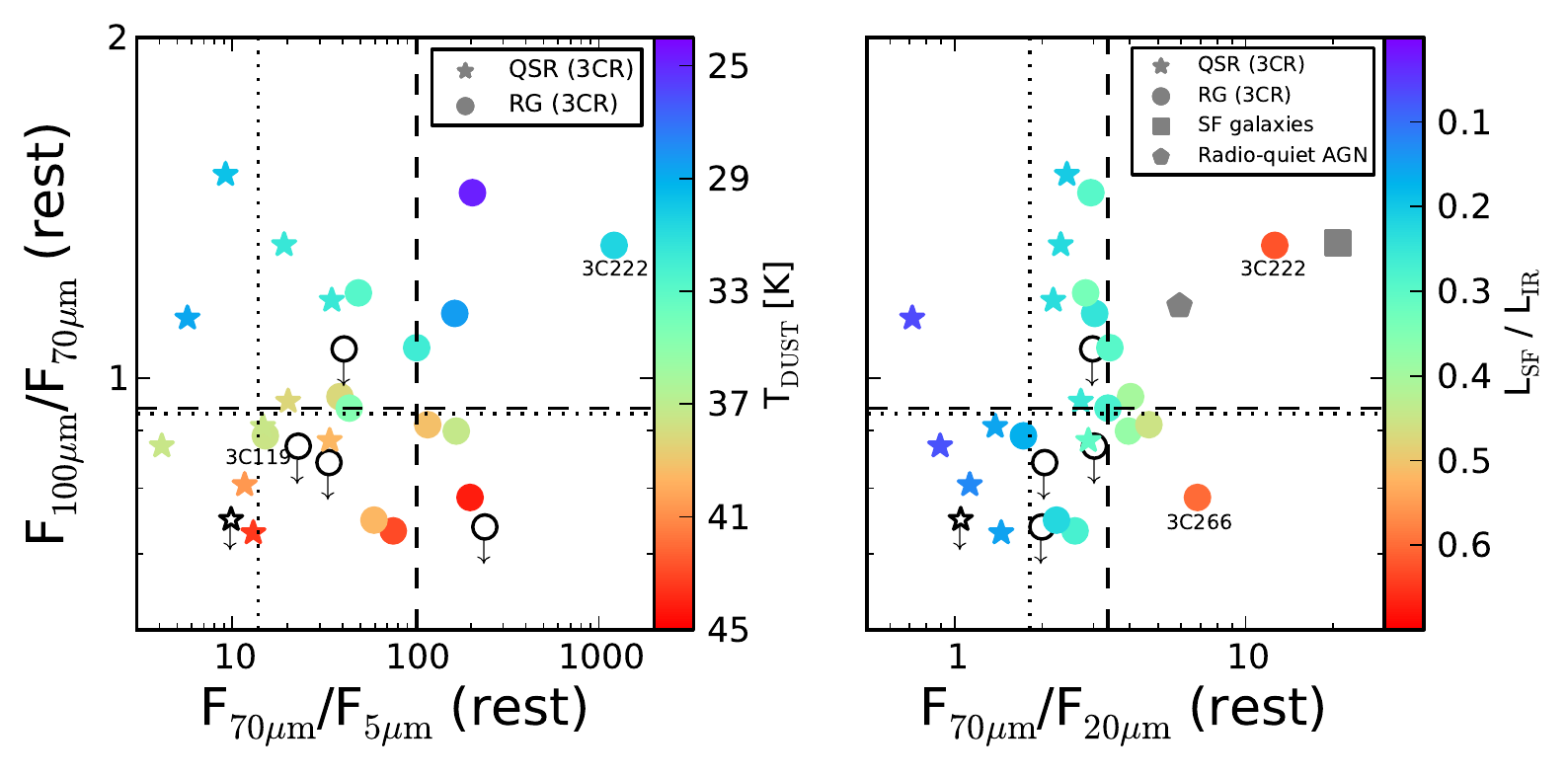}
 \caption{Rest-frame infrared color-color diagrams computed by fitting a multi-component model to the 
          observed spectral energy distributions of the 3CR objects. 
          The ordinate represents F$_{100\mu \mathrm{m}}$/F$_{70\mu \mathrm{m}}$ color in both plots, the abscissa the 
          F$_{70\mu \mathrm{m}}$/F$_{5\mu \mathrm{m}}$ (left plot) and F$_{70\mu \mathrm{m}}$/F$_{20\mu \mathrm{m}}$ 
          color (right plot). The 3CR quasars are shown with stars, the 3CR radio galaxies with circles. 
          Filled colored symbols represent objects detected in at least three \textit{Herschel} bands, whereas empty 
          symbols represent objects detected in only the two shortest \textit{Herschel} bands, for which 
          F$_{100\mu \mathrm{m}}$ is estimated as an upper limit. Dotted and dashed lines mark the corresponding median 
          colors of quasars and radio galaxies, respectively. In the left plot, the 3CR objects are color-coded according 
          to their cold dust temperatures, T$_\mathrm{DUST}$, as found from the SED-fitting. The T$_\mathrm{DUST}$ 
          scale is indicated in the left vertical color bar. In the right plot, the 3CR objects are color-coded according to 
          their fractional star formation luminosity, L$_\mathrm{SF}$/L$_\mathrm{IR}$. The 
          L$_\mathrm{SF}$/L$_\mathrm{IR}$ scale is indicated in the right vertical color bar. T$_\mathrm{DUST}$ and 
          L$_\mathrm{SF}$/L$_\mathrm{IR}$ are listed in Table~\ref{fig1}. Grey symbols show the median colors of the 
          comparison star-forming galaxies at $z\sim2$ (square) and radio-quiet AGN (pentagon), both from \citet{Kirkpatrick12}.}
 \label{fig1}
\end{figure}

\begin{figure}
 \epsscale{1.0} 
 \plotone{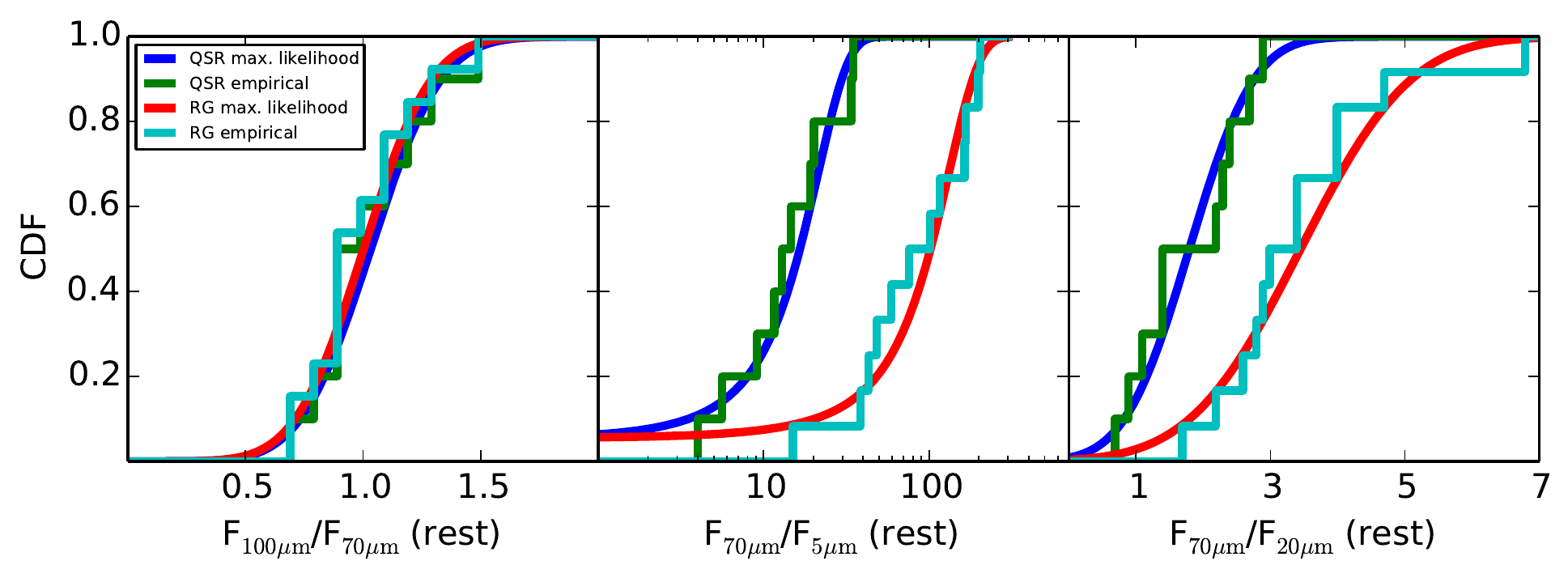}
 \caption{Maximum likelihood and empirical (measured) cumulative distribution functions of 
          F$_{100\mu \mathrm{m}}$/F$_{70\mu \mathrm{m}}$ (left), 
          F$_{70\mu \mathrm{m}}$/F$_{5\mu \mathrm{m}}$ (middle), and 
          F$_{70\mu \mathrm{m}}$/F$_{20\mu \mathrm{m}}$ (right) colors of the 3CR objects. 
          Quasars (blue/green) and radio galaxies (red/cyan) have different 
          F$_{70\mu \mathrm{m}}$/F$_{5\mu \mathrm{m}}$ and F$_{70\mu \mathrm{m}}$/F$_{20\mu \mathrm{m}}$ 
          distributions and indistinguishable F$_{100\mu \mathrm{m}}$/F$_{70\mu \mathrm{m}}$ distributions, 
          both in line with expectations from the unification model of radio-loud AGN. 
          (Left) Parameters of quasar maximum likelihood distribution: $\mu$=1.0, $\sigma$=0.2. 
          Parameters of radio galaxy maximum likelihood distribution: $\mu$=1.0, $\sigma$=0.2.
          (Middle) Parameters of quasar maximum likelihood distribution: $\mu$=16.6, $\sigma$=10.2. 
          Parameters of radio galaxy maximum likelihood distribution: $\mu$=102.1, $\sigma$=63.9.
          (Right) Parameters of quasar maximum likelihood distribution: $\mu$=1.8, $\sigma$=0.7. 
          Parameters of radio galaxy maximum likelihood distribution: $\mu$=3.4, $\sigma$=1.3.
          3C~222 has not been included in the F$_{70\mu \mathrm{m}}$/F$_{5\mu \mathrm{m}}$ and 
          F$_{70\mu \mathrm{m}}$/F$_{20\mu \mathrm{m}}$ distributions of radio galaxies (see text).
          }
 \label{fig2}
\end{figure}

In Fig.~\ref{fig1} we show infrared color-color diagrams for this sample of 13 RGs and 10 QSRs. The 
distributions of F$_{100\mu \mathrm{m}}$/F$_{70\mu \mathrm{m}}$\footnote{monochromatic flux densities 
are given in units of erg s$^{-1}$ cm$^{-2}$ Hz$^{-1}$} color for RGs and QSRs in Fig.~\ref{fig1} are 
nearly identical, with median values 0.9. The color coding in Fig.~\ref{fig1} (left) demonstrates that 
the factor of $\sim$2 scatter around the median FIR colors results from differences in the cold 
dust temperatures of individual objects, which range from $\sim25$ to $\sim45$~K \citep{Podigachoski15}. 
The emission from hosts with dust temperatures $<$ 33~K peaks at wavelengths $>$ 70~$\mu$m resulting in 
F$_{100\mu \mathrm{m}}$/F$_{70\mu \mathrm{m}}$ $>$ 1. 
We apply both maximum likelihood and empirical methods to statistically compare the measured RG 
and QSR F$_{100\mu \mathrm{m}}$/F$_{70\mu \mathrm{m}}$ colors. Assuming that the colors are normally 
distributed, we compute the parameters of the distributions which maximize the probability of measuring 
such colors. The means and standard deviations of these maximum likelihood distributions are 1.0 and 0.2, 
respectively, for both the RGs and QSRs. The corresponding maximum likelihood cumulative distribution 
functions and the empirical (measured) cumulative distributions are presented in Fig.~\ref{fig2} (left). 
The Kolmogorov-Smirnov (KS) test reveals that the two samples are drawn from the same distribution 
(KS$_{\mathrm{statistic}}$=0.07, p=1). We conclude that the FIR colors of these 3CR RGs and QSRs are 
indistinguishable (i.e., orientation invariant), as expected in the unification model for AGN host galaxies 
with similar star formation rates.  

In contrast, the RGs and QSRs show clear separation in their 
F$_{70\mu \mathrm{m}}$/F$_{5\mu \mathrm{m}}$ colors (Fig.~\ref{fig1}, left): their median colors are 101.4 
and 13.9, respectively. One RG, 3C~119, has colors which overlap the QSRs in Fig.~\ref{fig1} (left), not 
surprisingly, because 3C~119 is a well-known \citep{DeVries97} borderline source where some of the hot dust 
is visible, and the UV/visible AGN continuum is only partly hidden. The 5~$\mu$m emission of another 
RG, 3C~222, is only weakly constrained \citep{Podigachoski15}, resulting in an unusually red 
F$_{70\mu \mathrm{m}}$/F$_{5\mu \mathrm{m}}$ color for this object. Considering this particular value as an 
upper limit, we obtain KS$_{\mathrm{statistic}}$=0.92 (p=5$\times$10$^{-5}$), and maximum likelihood 
distributions as plotted in Fig.~\ref{fig2} (middle). These clearly different distributions are driven by the 
differences between the RG and QSR SEDs at 5~$\mu$m. RGs are fainter than QSRs at 5~$\mu$m, as expected if 
their hot inner torus dust emission is obscured due to its edge-on viewing angle \citep{Haas08,Leipski10,Dicken14}.  
Our results support the hypothesis that F$_{70\mu \mathrm{m}}$/F$_{5\mu \mathrm{m}}$ to a large 
extent reveals the level of obscuration in powerful radio-loud objects, and as such is an orientation indicator 
for that AGN population. However, intrinsic differences in this ratio for individual objects are not ruled out, 
and may contribute to the observed scatter.     

Although our adopted SED-fitting approach is physically grounded and provides us 
with a close approximation of the objects' infrared SEDs, we further explore the 
robustness of the results presented above without using dust emission models. 
To this end, we employ a simple (model-independent) linear interpolation of the 
observed flux values to obtain the rest-frame colors and create the color-color 
diagrams. We apply this method to the sample of objects detected in at least three 
\textit{Herschel} bands, and find that the RGs and QSRs remain clearly separated 
in the F$_{70\mu \mathrm{m}}$/F$_{5\mu \mathrm{m}}$ color, with median values 
of 83.9 and 12.3, respectively. Their respective median 
F$_{100\mu \mathrm{m}}$/F$_{70\mu \mathrm{m}}$ colors (1.0 and 0.9) remain 
indistinguishable, similarly to the results obtained when following the SED-fitting approach. 

In addition to the 23 3CR hosts detected in at least three \textit{Herschel} bands, 
seven hosts are detected in only the two PACS bands while remaining below the detection 
limit in the SPIRE bands. Given the low redshift range ($1<z\leq1.35$) of five of these 
seven hosts, the PACS 160~$\mu$m band probes their cold dust emission at rest-frame 
$\sim$ 70~$\mu$m allowing their inclusion (empty symbols) in the color-color diagrams 
presented in Fig.~\ref{fig1}. While this inclusion does not change the trends discussed above, 
their F$_{70\mu \mathrm{m}}$/F$_{5\mu \mathrm{m}}$ colors tend to be somewhat bluer than 
the median colors of the 23 objects detected in at least three \textit{Herschel} bands. 
This is likely a consequence of their lower levels of star formation \citep{Podigachoski15}. 
As such, it is important to note that the present FIR test of 
unification can directly be applied only to the long wavelength detected, i.e., the 
star-forming objects from the high-$z$ 3CR sample. The 31 remaining hosts from the 
sample, which are either detected only in the PACS 70~$\mu$m band or not detected at all, 
have weak or absent star formation and thus no definitive conclusions can be drawn. 
Nevertheless, the non-detection fractions at SPIRE 250~$\mu$m for RGs and QSRs are 
comparable ($\sim65\%$ and $\sim58\%$ respectively), and thus do not represent 
evidence against the unification model. 

The high level of obscuration in the MIR prevents the 
F$_{70\mu \mathrm{m}}$/F$_{5\mu \mathrm{m}}$ color from directly probing the relative 
importance of the SF and AGN activity. We therefore use the AGN-powered circumnuclear 
dust emission at 20~$\mu$m \citep[near the peak emission from the torus, e.g.,][]{Hoenig&Kishimoto10}, 
and employ the F$_{70\mu \mathrm{m}}$/F$_{20\mu \mathrm{m}}$ color to probe the dominant power 
source in the infrared. As shown in Fig.~\ref{fig1} (right), the 
F$_{70\mu \mathrm{m}}$/F$_{20\mu \mathrm{m}}$ RG and QSR colors are closer to each 
other than the F$_{70\mu \mathrm{m}}$/F$_{5\mu \mathrm{m}}$ colors, but still have an 
offset, with median values of 3.4 and 1.8 respectively (KS$_{\mathrm{statistic}}$=0.65, p=0.01; 
see Fig.~\ref{fig2}, right), confirming that the level of obscuration at 20~$\mu$m is substantially 
lower than that at 5~$\mu$m. The symbols in Fig.~\ref{fig1} (right) are 
color-coded according to the fractional star formation luminosity, 
L$_\mathrm{SF}$/L$_\mathrm{IR}$, as inferred in \citet{Podigachoski15}. 
This color-coding demonstrates that, despite the strong star 
formation activity of several hundreds of solar masses per year, the total infrared 
luminosity of the 3CR hosts is generally dominated by their AGN activity, regardless 
if that activity is obscured or not \citep{Podigachoski15}. The only exceptions are 
two RGs, 3C~266 and 3C~222, which have similar L$_\mathrm{SF}$ but lower IR AGN 
luminosities than the other RGs \citep{Podigachoski15}. However, a general increase 
in the L$_\mathrm{SF}$/L$_\mathrm{IR}$ ratio of 3CR objects, from $\sim10\%$ 
to $\sim60\%$, is observed when progressing to larger 
F$_{70\mu \mathrm{m}}$/F$_{20\mu \mathrm{m}}$ values, i.e., to redder colors. 

We examine this trend further by comparing the 3CR AGN infrared colors to those of 
typical star-forming (SF) galaxies where the infrared emission draws uniquely from 
star formation. The SFR strongly correlates with stellar mass and redshift 
\citep{Elbaz07}; we therefore select $z\sim2$ SF galaxies from \citet{Kirkpatrick12} 
with stellar masses and redshifts comparable to those of the 3CR hosts \citep[which have 
M$_{\mathrm{stellar}}$ $\gtrsim$ 10$^{11}$ M$_{\odot}$, e.g.,][]{Best98,DeBreuck10}. 
Given that stars in these SF galaxies are formed at rates \citep{Kirkpatrick12} comparable to 
those in the 3CR hosts, the L$_\mathrm{SF}$/L$_\mathrm{IR}$$\sim$1 of the SF galaxies 
are consistent with their red F$_{70\mu \mathrm{m}}$/F$_{20\mu \mathrm{m}}$ colors (Fig.~\ref{fig1} right). 
The bluer F$_{70\mu \mathrm{m}}$/F$_{20\mu \mathrm{m}}$ colors of the powerful 3CR 
AGN are likely a consequence of the strong torus emission at 20~$\mu$m. The average 
cold dust temperature of the SF galaxies is 28$\pm$2~K \citep{Kirkpatrick12}, hence their 
FIR colors are also consistent with the temperature trends found for the 3CR hosts. 
The F$_{70\mu \mathrm{m}}$/F$_{20\mu \mathrm{m}}$ colors of powerful AGN are offset from 
those of SF galaxies, and hence, in addition to being indicators of orientation, FIR/MIR 
colors also distinguish between active and SF galaxies in the most luminous star-forming 
objects, analogous to the case for local Seyfert galaxies \citep[e.g.,][]{DeGrijp85}.

A limitation for the general unification discussion is the fact that radio-loud AGN, such as 
the 3CR sources investigated here, represent a small ($\sim10\%$) fraction of the complete 
AGN population. It is therefore important to compare the colors of the powerful 3CR 
radio-loud AGN to those of powerful radio-quiet AGN. To achieve this, we select a sample 
of distant ($z_{\mathrm{med}}\sim$ 1.9), luminous, radio-quiet AGN 
(L$_\mathrm{2-10 keV}$ $\sim$ 5$\times$10$^{44}$ erg s$^{-1}$) hosted by massive 
(M$_{\mathrm{stellar}}$ $\sim$ 10$^{11}$ M$_{\odot}$) star-forming galaxies, from 
\citet{Kirkpatrick12}. As shown in Fig.~\ref{fig1} (right) the radio-quiet AGN have 
F$_{70\mu \mathrm{m}}$/F$_{20\mu \mathrm{m}}$ colors redder than those of the 3CR RGs, but 
bluer than those of the SF galaxies. Given that the SF activity in these radio-quiet AGN 
accounts for 56\% of the total infrared luminosity \citep{Kirkpatrick12}, this finding agrees 
with the trend of increasing L$_\mathrm{SF}$/L$_\mathrm{IR}$ towards larger 
F$_{70\mu \mathrm{m}}$/F$_{20\mu \mathrm{m}}$ values for the 3CR objects discussed above. 
\section{Conclusion}
In conclusion, far- to mid-infrared colors represent a useful diagnostic of the orientation 
of powerful radio-loud AGN, and support the unification model for these objects. At 
wavelengths longer than 20~$\mu$m, where obscuration is a small effect, these colors are also 
indicators of the relative contributions of star formation and nuclear activity in galaxies. 
These diagnostics may prove useful in future studies of large samples of dusty star-forming 
galaxies/AGN to investigate the duty cycle of nuclear activity, feedback, and AGN and galaxy 
growth. 
\acknowledgments
We thank the referee, Patrick Ogle, for a very constructive report which improved 
the manuscript. 
Data were taken from the \textit{Herschel} Guaranteed Time project \textit{The 
Herschel Legacy of distant radio-loud AGN} (PI: PB). PP acknowledges the Nederlandse 
Organisatie voor Wetenschappelijk Onderzoek (NWO) for a PhD fellowship. 
The \textit{Herschel} spacecraft was designed, built, tested, and launched under a 
contract to ESA managed by the \textit{Herschel}/\textit{Planck} Project team by an industrial 
consortium under the overall responsibility of the prime contractor Thales 
Alenia Space (Cannes), and including Astrium (Friedrichshafen) responsible 
for the payload module and for system testing at spacecraft level, Thales 
Alenia Space (Turin) responsible for the service module, and Astrium (Toulouse) 
responsible for the telescope, with in excess of a hundred subcontractors.
PACS has been developed by a consortium of institutes led by MPE 
(Germany) and including UVIE (Austria); KU Leuven, CSL, IMEC 
(Belgium); CEA, LAM (France); MPIA (Germany); INAF-IFSI/OAA/OAP/OAT, 
LENS, SISSA (Italy); IAC (Spain). This development has been supported 
by the funding agencies BMVIT (Austria), ESA-PRODEX (Belgium), 
CEA/CNES (France), DLR (Germany), ASI/INAF (Italy), and CICYT/MCYT (Spain).
SPIRE has been developed by a consortium of institutes led 
by Cardiff University (UK) and including Univ. Lethbridge (Canada); 
NAOC (China); CEA, LAM (France); IFSI, Univ. Padua (Italy); IAC 
(Spain); Stockholm Observatory (Sweden); Imperial College London, RAL, 
UCL-MSSL, UKATC, Univ. Sussex (UK); and Caltech, JPL, NHSC, Univ. 
Colorado (USA). This development has been supported by national 
funding agencies: CSA (Canada); NAOC (China); CEA, CNES, CNRS 
(France); ASI (Italy); MCINN (Spain); SNSB (Sweden); STFC, UKSA (UK); 
and NASA (USA).
This work is partly based on observations made with the \textit{Spitzer} Space 
Telescope, which is operated by the Jet Propulsion Laboratory, 
California Institute of Technology under a contract with NASA.

{\it Facilities:} \facility{Herschel}, \facility{Spitzer}.

\clearpage

\clearpage

\clearpage

\begin{deluxetable}{lccccccc}
\tablewidth{0pt}
\tabletypesize{\scriptsize}
\tablecaption{Properties of the 3CR objects plotted in Fig.~\ref{fig1}. 
              The top 23 objects are detected in at least three \textit{Herschel} bands, 
              whereas the bottom 5 objects in only two \textit{Herschel} bands. The redshifts 
              of the objects are taken from \citet{Spinrad85}. Columns 4 through 8 are 
              computed by fitting the full infrared spectral energy distributions of the hosts 
              as described in the text. 
              }
\tablehead{
\colhead{Name} & \colhead{Type} & \colhead{$z$} & \colhead{F$_{70\mu \mathrm{m}}$/F$_{5\mu \mathrm{m}}$} &
\colhead{F$_{70\mu \mathrm{m}}$/F$_{20\mu \mathrm{m}}$} & \colhead{F$_{100\mu \mathrm{m}}$/F$_{70\mu \mathrm{m}}$} &  
\colhead{T$_\mathrm{DUST}$ [K]} & \colhead{L$_\mathrm{SF}$/L$_\mathrm{IR}$}}
\startdata
3C~002    & QSR & 1.04 & 19.2   & 2.3  & 1.3 & 31.6 & 0.23 \\
3C~014    & QSR & 1.47 & 4.1    & 0.9  & 0.9 & 37.5 & 0.08 \\
3C~068.2  & RG  & 1.57 & 163.7  & 3.0  & 1.1 & 28.5 & 0.25 \\ 
3C~119    & RG  & 1.02 & 15.1   & 1.7  & 0.9 & 37.6 & 0.17 \\ 
3C~124    & RG  & 1.08 & 101.4  & 3.4  & 1.1 & 31.9 & 0.30 \\
3C~190    & QSR & 1.20 & 34.9   & 2.2  & 1.2 & 31.7 & 0.23 \\
3C~205    & QSR & 1.53 & 11.7   & 1.1  & 0.8 & 40.7 & 0.13 \\
3C~222    & RG  & 1.34 & 1212.7 & 12.6 & 1.3 & 30.6 & 0.62 \\
3C~245    & QSR & 1.03 & 5.7    & 0.7  & 1.1 & 28.8 & 0.07 \\
3C~256    & RG  & 1.82 & 75.6   & 2.6  & 0.7 & 43.0 & 0.27 \\
3C~257    & RG  & 2.47 & 38.6   & 4.0  & 1.0 & 38.2 & 0.40 \\
3C~266    & RG  & 1.27 & 198.3  & 6.8  & 0.8 & 44.2 & 0.60 \\
3C~270.1  & QSR & 1.52 & 13.0   & 1.4  & 0.7 & 43.5 & 0.15 \\
3C~297    & RG  & 1.41 & 204.5  & 2.9  & 1.5 & 24.9 & 0.30 \\
3C~298    & QSR & 1.44 & 14.7   & 1.4  & 0.9 & 37.5 & 0.16 \\
3C~305.1  & RG  & 1.13 & 43.4   & 3.4  & 0.9 & 34.7 & 0.28 \\
3C~318    & QSR & 1.57 & 34.0   & 2.9  & 0.9 & 39.6 & 0.31 \\
3C~324    & RG  & 1.21 & 59.3   & 2.2  & 0.7 & 39.6 & 0.23 \\
3C~368    & RG  & 1.13 & 166.8  & 4.0  & 0.9 & 37.3 & 0.38 \\
3C~432    & QSR & 1.80 & 9.2    & 2.4  & 1.5 & 29.8 & 0.21 \\
3C~454.0  & QSR & 1.76 & 20.2   & 2.7  & 1.0 & 38.3 & 0.26 \\
3C~454.1  & RG  & 1.84 & 116.4  & 4.7  & 0.9 & 39.3 & 0.45 \\
3C~470    & RG  & 1.65 & 48.4   & 2.8  & 1.2 & 32.8 & 0.34 \\
\cutinhead{Objects detected in only two \textit{Herschel} bands}
3C~013    & RG  & 1.35 & 238.8  & 2.0  & $<$0.7 &     &      \\
3C~210    & RG  & 1.17 & 40.7   & 3.0  & $<$1.1 &     &      \\
3C~220.2  & QSR & 1.16 & 9.8    & 1.0  & $<$0.8 &     &      \\
3C~356    & RG  & 1.08 & 33.7   & 2.0  & $<$0.8 &     &      \\
3C~469.1  & RG  & 1.34 & 22.9   & 3.0  & $<$0.9 &     &      \\
\enddata
\label{tab1}
\end{deluxetable}


\begin{thebibliography}{}
\bibitem[Antonucci(1993)]{Antonucci93} Antonucci, R.\ 1993, \araa, 31, 473 
\bibitem[Antonucci(2012)]{Antonucci12} Antonucci, R.\ 2012, Astronomical and Astrophysical Transactions, 27, 557 
\bibitem[Barthel(1989)]{Barthel89} Barthel, P.~D.\ 1989, \apj, 336, 606
\bibitem[Barthel et al.(2012)]{Barthel12} Barthel, P., Haas, M., Leipski, C., \& Wilkes, B.\ 2012, \apjl, 757, LL26
\bibitem[Best et al.(1998)]{Best98} Best, P.~N., Longair, M.~S., \& Roettgering, H.~J.~A.\ 1998, \mnras, 295, 549
\bibitem[Best \& Heckman(2012)]{Best&Heckman12} Best, P.~N., \& Heckman, T.~M.\ 2012, \mnras, 421, 1569 
\bibitem[Cleary et al.(2007)]{Cleary07} Cleary, K., Lawrence, C.~R., Marshall, J.~A., Hao, L., \& Meier, D.\ 2007, \apj, 660, 117
\bibitem[De Breuck et al.(2010)]{DeBreuck10} De Breuck, C., Seymour, N., Stern, D., et al.\ 2010, \apj, 725, 36
\bibitem[de Grijp et al.(1985)]{DeGrijp85} de Grijp, M.~H.~K., Miley, G.~K., Lub, J., \& de Jong, T.\ 1985, \nat, 314, 240
\bibitem[de Vries et al.(1997)]{DeVries97} de Vries, W.~H., O'Dea, C.~P., Baum, S.~A., et al.\ 1997, \apjs, 110, 191
\bibitem[Dicken et al.(2014)]{Dicken14} Dicken, D., Tadhunter, C., Morganti, R., et al.\ 2014, \apj, 788, 98 
\bibitem[Donley et al.(2012)]{Donley12} Donley, J.~L., Koekemoer, A.~M., Brusa, M., et al.\ 2012, \apj, 748, 142
\bibitem[Drouart et al.(2014)]{Drouart14} Drouart, G., De Breuck, C., Vernet, J., et al.\ 2014, \aap, 566, AA53 
\bibitem[Elbaz et al.(2007)]{Elbaz07} Elbaz, D., Daddi, E., Le Borgne, D., et al.\ 2007, \aap, 468, 33 
\bibitem[Griffin et al.(2010)]{Griffin10} Griffin, M.~J., Abergel, A., Abreu, A., et al.\ 2010, \aap, 518, LL3 
\bibitem[Haas et al.(2004)]{Haas04} Haas, M., M{\"u}ller, S.~A.~H., Bertoldi, F., et al.\ 2004, \aap, 424, 531 
\bibitem[Haas et al.(2008)]{Haas08} Haas, M., Willner, S.~P., Heymann, F., et al.\ 2008, \apj, 688, 122
\bibitem[Hes et al.(1995)]{Hes95} Hes, R., Barthel, P.~D., \& Hoekstra, H.\ 1995, \aap, 303, 8 
\bibitem[H{\"o}nig \& Kishimoto(2010)]{Hoenig&Kishimoto10} H{\"o}nig, S.~F., \& Kishimoto, M.\ 2010, \aap, 523, AA27 
\bibitem[Kirkpatrick et al.(2012)]{Kirkpatrick12} Kirkpatrick, A., Pope, A., Alexander, D.~M., et al.\ 2012, \apj, 759, 139
\bibitem[Kirkpatrick et al.(2013)]{Kirkpatrick13} Kirkpatrick, A., Pope, A., Charmandaris, V., et al.\ 2013, \apj, 763, 123
\bibitem[Lacy et al.(2004)]{Lacy04} Lacy, M., Storrie-Lombardi, L.~J., Sajina, A., et al.\ 2004, \apjs, 154, 166 
\bibitem[Leipski et al.(2010)]{Leipski10} Leipski, C., Haas, M., Willner, S.~P., et al.\ 2010, \apj, 717, 766
\bibitem[Meisenheimer et al.(2001)]{Meisenheimer01} Meisenheimer, K., Haas, M., M{\"u}ller, S.~A.~H., et al.\ 2001, \aap, 372, 719 
\bibitem[Ogle et al.(2006)]{Ogle06} Ogle, P., Whysong, D., \& Antonucci, R.\ 2006, \apj, 647, 161
\bibitem[Ogle et al.(2012)]{Ogle12} Ogle, P., Davies, J.~E., Appleton, P.~N., et al.\ 2012, \apj, 751, 13 
\bibitem[Pilbratt et al.(2010)]{Pilbratt10} Pilbratt, G.~L., Riedinger, J.~R., Passvogel, T., et al.\ 2010, \aap, 518, LL1 
\bibitem[Podigachoski et al.(2015)]{Podigachoski15} Podigachoski, P., Barthel, P.~D., Haas, M., et al.\ 2015, \aap, 575, AA80 
\bibitem[Poglitsch et al.(2010)]{Poglitsch10} Poglitsch, A., Waelkens, C., Geis, N., et al.\ 2010, \aap, 518, LL2 
\bibitem[Rowan-Robinson(1995)]{Rowan-Robinson95} Rowan-Robinson, M.\ 1995, \mnras, 272, 737
\bibitem[Schweitzer et al.(2006)]{Schweitzer06} Schweitzer, M., Lutz, D., Sturm, E., et al.\ 2006, \apj, 649, 79
\bibitem[Singal(1993)]{Singal93} Singal, A.~K.\ 1993, \mnras, 262, L27 
\bibitem[Spinrad et al.(1985)]{Spinrad85} Spinrad, H., Marr, J., Aguilar, L., \& Djorgovski, S.\ 1985, \pasp, 97, 932
\bibitem[Stern et al.(2005)]{Stern05} Stern, D., Eisenhardt, P., Gorjian, V., et al.\ 2005, \apj, 631, 163
\bibitem[Tadhunter et al.(2014)]{Tadhunter14} Tadhunter, C., Dicken, D., Morganti, R., et al.\ 2014, \mnras, 445, L51 
\bibitem[Urry \& Padovani(1995)]{Urry&Padovani95} Urry, C.~M., \& Padovani, P.\ 1995, \pasp, 107, 803 
\bibitem[van Bemmel et al.(2000)]{vanBemmel00} van Bemmel, I.~M., Barthel, P.~D., \& de Graauw, T.\ 2000, \aap, 359, 523 
\bibitem[Werner et al.(2004)]{Werner04} Werner, M.~W., Roellig, T.~L., Low, F.~J., et al.\ 2004, \apjs, 154, 1
\bibitem[Wilkes et al.(2013)]{Wilkes13} Wilkes, B.~J., Kuraszkiewicz, J., Haas, M., et al.\ 2013, \apj, 773, 15
\end{thebibliography}
\end{document}